\documentclass[twocolumn]{revtex4}

\pdfoutput=1
\usepackage{hyperref}
\usepackage{bm}

\usepackage{amsmath}
\usepackage{amssymb}
\usepackage{amsthm}
\usepackage{array}
\usepackage{xy}

\usepackage{graphicx}
\usepackage{subfigure}
\usepackage{epstopdf}
\usepackage{grffile}

\usepackage{titlesec}
\usepackage{lipsum}
\usepackage{tipa}

\usepackage{booktabs}
\usepackage{csvsimple}

\newcommand{\hrmsd}{0.179}
\newcommand{\heavyrmsd}{0.051}

\begin{document}

%\title{Coarse-grained protein model with oriented sites for accurate and efficient coarse-grained or multiscale simulations}
%\title{High-resolution coarse-grained modeling with few anisotropic sites}
\title{High-resolution coarse-grained modeling using oriented coarse-grained sites}

%Accurate coarse-grained modeling of hierarchical macromolecular assemblies}
%Group meeting, user meeting: A coarse-grained model for the hierarchical assembly of peptoid nanosheets
%MRS: Modeling the nonequilibrium assembly of sequence-specific polymers into complex structures with a novel coarse-grained modeling approach
%MRS spring: Hierarchical assembly of peptoid nanosheets catalyzed by an air-water interface
%APS: Hierarchical assembly of peptoid nanosheets catalyzed by an air-water interface

\author{Thomas K. Haxton}

\affiliation{Molecular Foundry, Lawrence Berkeley National Laboratory, Berkeley, CA 94720, USA}

\begin{abstract}
We introduce a method to bring nearly atomistic resolution to coarse-grained models, and we apply the method to proteins.  Using a small number of coarse-grained sites (about one per eight atoms) but assigning an \textit{independent three-dimensional orientation} to each site, we preferentially integrate out stiff degrees of freedom (bond lengths and angles, as well as dihedral angles in rings) that are accurately approximated by their average values, while retaining soft degrees of freedom (unconstrained dihedral angles) mostly responsible for conformational variability.  We demonstrate that our scheme retains nearly atomistic resolution by mapping all experimental protein configurations in the Protein Data Bank onto coarse-grained configurations, then analytically backmapping those configurations back to all-atom configurations.  This roundtrip mapping throws away all information associated with the eliminated (stiff) degrees of freedom except for their average values, which we use to construct optimal backmapping functions.  Despite the 4:1 reduction in the number of degrees of freedom, we find that heavy atoms move only \heavyrmsd~\AA~on average during the roundtrip mapping, while hydrogens move \hrmsd~\AA~on average, an unprecedented combination of efficiency and accuracy among coarse-grained protein models.  We discuss the advantages of such a high-resolution model for parameterizing effective interactions and accurately calculating observables through direct or multiscale simulations.
%As long as computationally tractable effective interactions can be found that accurately approximate the many-body coarse-grained potential of mean force, our results indicate that our model could reproduce all observables of the all-atom system, even those depending explicitly on atomic positions, with errors proportional to these root-mean-square displacements.  We discuss the advantages of using oriented coarse-grained sites for directly computing accurate effective interactions without fitting, and we discuss the advantages of high-resolution backmapped atomic coordinates for incorporating our model into various multiscale simulation schemes.
\end{abstract}
%600 characters

%\date{\today}

%\pacs{}

\maketitle

%Many of the most important processes in molecular biology, including allostery~\cite{Collier2013, Cui2008}, enzyme catalysis~\cite{McGeagh2011}, molecular recognition~\cite{Tuffery2012}, protein homeostasis~\cite{England2008}, and nucleic acid metabolism~\cite{Wang2013}, involve the cooperative motion of large but precisely self-assembled~\cite{Bowman2011} biomolecules.
%Molecular simulation--especially when coupled closely with experiment--can provide a microscopic understanding of these processes, leading to a better understanding of biological function and treatment of diseases.

The development of accurate classical force-fields and special-purpose supercomputers have allowed all-atom molecular simulations to provide a rich microscopic view of many complex processes in molecular biology.
%mostly about sampling~\cite{Zuckerman2011}
For example, all-atom molecular dynamics simulations are now able to explore in atomic detail the folding kinetics of fast-folding protein domains~\cite{Shaw2010, Lindorff-Larsen2011}, engineered protein dimers~\cite{Piana2013}, and small, naturally-occuring proteins~\cite{Piana2012b}, using simulations spanning milliseconds for proteins containing up to 100 residues.  
%35-residue domain~\cite{Shaw2010}
%10 to 80 residue domains, fold in 0.6 to 65 microsecond~\cite{Lindorff-Larsen2011}
%~\cite{Piana2012b} ubiquitin 76-residue; experimentally ms time scale
%~\cite{Piana2013} aggregate 4 ms 114-residue designed homodimer, about 0.3 ms folding time

While hardware and software advances will continue to advance the dynamic range of all-atom simulations, accessible time and length scales will always be limited by the short time steps (typically $10^{-15}$ s) required to resolve atomic vibrational motion~\cite{Feenstra1999} and the computational expense of propagating all the atomic degrees of freedom.  Thus, many important biological processes remain significantly beyond the frontier of all-atom simulations.  For example, understanding complex and competing assembly pathways of soluble and insoluble aggregates is crucial for developing therapies for Alzheimer's disease, but these pathways can each involve ten or more 40-residue peptides and span hours or days~\cite{Necula2007, Bartolini2011, Nguyen2014}. 

Coarse-grained modeling presents an appealing way to extend the time and length scales accessible to molecular simulations~\cite{Nielson2004, Clementi2008, Murtola2009, Tozzini2010, Trylska2010, Kamerlin2011, Hyeon2011, Takada2012, Shinoda2012, Saunders2013, Noid2013, Brini2013}.  By reducing the number of degrees of freedom by a factor of $N$, coarse-grained simulations decrease the number of short-range force calculations and increase accessible time scales both by factors of $N^2$.   Averaging out the stiffest degrees of freedom allows longer time steps in molecular dynamics simulations (or increased trial step sizes in Monte Carlo simulations), further increasing the accessible dynamic range.  Coarse-grained models can be used by themselves or as components in multiscale simulation schemes~\cite{Tschop1998, Shih2007, Perlmutter2009, Neri2005, Shi2006, Machado2011, Mamonov2012, diPasquale2012, Leguebe2012, Lyman2006, Lyman2006b, Christen2006, Moritsugu2010} that use coarse-grained simulations to accelerate sampling or dynamics of related all-atom simulations.  

Information is lost when integrating out degrees of freedom, so the increased efficiency of coarse-grained modeling is always paid for with a decrease in accuracy~\cite{Louis2002, Stillinger2002, Johnson2007, Kowalczyk2009, Kowalczyk2011}.  Because complex biological systems are sensitive to decreases in accuracy, coarse-grained modeling has in many cases failed in a qualitative way to extend the reach of molecular simulation.  For example, unbiased coarse-grained models have not yet successfully folded any proteins that cannot be folded by all-atom simulations.  Thus, methods to improve the accuracy of coarse-grained models are urgently needed.
%However, it has been shown that the accuracy of isotropic coarse-grained models declines as the underlying all-atom system becomes more anisotropic, due to the fact that more information is lost when averaging spherically over anisotropic interactions than over isotropic ones~\cite{Kowalczyk2009, Kowalczyk2011}.

The accuracy of coarse-grained models can be improved by optimizing one or both of the two aspects of coarse-grained modeling: (1) the mapping between all-atom and coarse-grained degrees of freedom and (2) the effective interactions among the coarse-grained degrees of freedom.  Many authors have focused on the second aspect, developing efficient methods to parameterize interactions so that differences between coarse-grained simulations and reference all-atom simulations are minimized.  Examples include iterative Boltzmann inversion that matches pair distribution functions~\cite{Reith2003}, the multiscale coarse-graining method that matches forces~\cite{Izvekov2005, Noid2008, Noid2008b, Izvekov2010}, and the relative entropy method that minimizes the relative entropy between coarse-grained and all-atom ensembles~\cite{Shell2008, Chaimovich2010, Chaimovich2011, Carmichael2012}.  

Some authors have also begun exploring different ways to map atomic positions onto coarse-grained sites, which in turn dictates the types of effective interactions available.  Most coarse-grained models are point-site models, mapping groups of atoms onto structureless coarse-grained sites.  In many point-site models, the effective interactions are analogous to all-atom forcefields: interactions between covalently connected sites are decomposed into bond stretching, bending, and twisting terms, and non-bonded interactions are spherically symmetric.  Although point-site models discard all information about the structure of each site's associated atoms, many authors have developed methods to estimate some aspects of a site's structure from the positions of neighboring sites, allowing the introduction of directional non-bonded terms to more accurately model directional interactions like hydrogen bonding~\cite{Liwo2004, Yap2008, Majek2009, Alemani2009, Enciso2010, Enciso2012, Kar2013, Liwo1997, Ouldridge2010, Linak2011, Sulc2013, Orsi2011}.  Some protein models have estimated the orientation of hydrogen-bonding peptide groups from the positions of contiguous $\alpha$-carbons~\cite{Liwo2004, Yap2008, Majek2009, Alemani2009, Enciso2010, Enciso2012} or from the positions of multiple backbone sites per residue~\cite{Kar2013}, while others have estimated the orientation of ellipsoidal side-chain sites using the position of the neighboring backbone site~\cite{Liwo1997}.  
Similar efforts have also been applied to DNA~\cite{Ouldridge2010, Linak2011, Sulc2013} and lipid~\cite{Orsi2011} models.

In one protein example, recognizing that the orientation of peptide bond dipoles in the Protein Data Bank (PDB)~\cite{PDB} correlates with the angle among the three nearest $\alpha$-carbons, Alemani \textit{et al.}~created a model defining the peptide bond dipole orientation as a function of the $\alpha$-carbon angle~\cite{Alemani2009}.
%in the frame defined by the three $\alpha$-carbons
By including a dipole interaction, a bistable bond angle interaction, and a term coupling consecutive dihedral angles along the $\alpha$-carbon chain, Alemani \textit{et al.}~could tune between alpha-helix and beta-sheet secondary structures.  Although this approach illustrates a minimal set of ingredients needed to select secondary structures, the low correlation in the PDB between peptide bond angles and backbone angles \textit{within} each class of secondary structure (Fig.~1 in Ref.~\cite{Alemani2009}) illustrates the resolution limit of such an approach.

A few authors have increased the resolution of coarse-grained models by introducing structure into individual coarse-grained sites in the form of vectors.  Morriss-Andrews \textit{et al.}~created a nucleic acid model with ellipsoidal nucleobases allowed to freely rotate about the vector between the base site and its associated backbone site~\cite{Morriss-Andrews2010}.  Spiga \textit{et al.}~created a protein model with electric dipoles at polar side-chain sites allowed to freely rotate in all directions~\cite{Spiga2013}.  Very recently, our group created a model for peptoids (positional isomers of peptides) assigning a rotatable vector to each site, using the vectors to construct orientation-dependent bonded and non-bonded interactions~\cite{Haxton2014peptoidmethods}.

Here, we introduce a new method to bring atomistic resolution to coarse-grained models: assign an \textit{independent three-dimensional orientation} to each site.  While adding orientations to a point-site model increases the number of degrees of freedom per site from three to six, it efficiently uses the separation of energy scales characterizing organic chemistry: bond lengths and angles are stiff, while dihedral angles (except in rings and double bonds) are soft.  Preferentially integrating out the stiff degrees of freedom allows a dramatic reduction in the number of degrees of freedom with a minimal loss of accuracy.  As we discuss in Section~\ref{mapping},
%the accuracy of a coarse-grained model can be separated into two components, 
the ability of a coarse-grained model to accurately calculate observables depends on two contributions to its accuracy:
%(1) the ability of effective interactions to approximate an exact but computationally intractable potential of mean force and (2) the ability of the model to reproduce an equilibrium distribution of atomic coordinates after a \textit{roundtrip mapping}.  
(1) the accuracy of the effective interactions used to approximate the many-body potential of mean force and
(2) the accuracy of atomic coordinates produced during a \textit{roundtrip mapping}.
In a roundtrip mapping, atomic coordinates are mapped onto a coarse-grained model (throwing out information), then 
%analytical backmapping functions are applied to the coarse-grained coordinates to produce a new set of atomic coordinates
a new set of atomic coordinates are produced from the coarse-grained coordinates using analytical backmapping functions~\cite{Gopal2009}.  
In Section~\ref{results}, we explicitly demonstrate that an oriented coarse-grained protein model with a 4:1 mapping (eight atoms per oriented site) can achieve an unprecedented combination of efficiency and accuracy in a roundtrip mapping.
%develop forward mapping and backmapping functions for an oriented protein model with a 4:1 mapping (eight atoms per oriented site).  
Using all experimental protein structures from the Protein Data Bank~\cite{PDB} as a large and experimentally based proxy for an equilibrium distribution of atomic coordinates, we optimize backmapping functions to reduce the information lost during a roundtrip mapping.  After this optimization, we find that heavy atoms move only \heavyrmsd~\AA~on average during the roundtrip mapping, while hydrogens move \hrmsd~\AA.
%This can be demonstrated by mapping all-atom configurations onto coarse-grained configurations, then analytically \textit{backmapping} the coarse-grained configurations back to all-atom configurations.  This roundtrip mapping throws away all information associated with the eliminated (stiff) degrees of freedom except for their average values, which can be used to construct optimal backmapping functions.  Applying this demonstration to all experimental protein configurations in the Protein Data Bank~\cite{PDB}, we show below that a protein model with a 4:1 mapping (eight atoms per oriented site) can be optimized such that heavy atoms move only \heavyrmsd~\AA~on average, while hydrogens move \hrmsd~\AA~on average.
%This extreme accuracy exceeds the previously most accurate coarse-grained model by a factor of two, 
% PRIMO: 9.9 pm average for heavy atoms
%despite containing only half as many degrees of freedom.  

Although the mapping between all-atom and coarse-grained representations is just one half of a coarse-grained model, our results bode well for developing tractable effective interactions that introduce a minimal amount of additional error.
%Although the mapping between all-atom and coarse-grained representations is just one half of a coarse-grained model, our results bode well for developing the effective interactions that would complete the model, 
%as discussed in Section~\ref{interactions}.  
%The high resolution of the backmapped configurations means that coarse-grained simulations using \textit{optimal} effective interactions could reproduce any observable measurable in the all-atom system with a similarly high resolution, even observables depending explicitly on atomic positions.  In general, coarse-grained models can only achieve this resolution limit exactly if many-body effective interactions of all orders are included.
%First, the accuracy of our backmapped configurations suggests that coarse-grained simulations using \textit{ideal} effective interactions could reproduce any observable measurable in the all-atom system with a similarly high resolution, even observables depending explicitly on atomic positions.  Second, the simplicity of our backmapping functions (see the next section) illustrates a straightforward approach to calculate effective interactions approaching these ideal ones.  
In contrast to point-site models, the position and orientation of a single site can predict the position of surrounding atoms associated with that site.  Thus, the atomistic interactions between two sites can be expressed directly as a potential of mean force, calculated \textit{without fitting} by recording the distribution of relative positions and orientations of the two sites in all-atom simulations.
Although approximate factorizations and symmetries may simplify some of these calculations, in general we expect that these potentials of mean force will be six-dimensional (three positional and three orientational degrees of freedom).  We leave the substantial computational task of calculating these potentials of mean force for future work.
%Moreover, although the use of orientation-dependent effective pair potentials should reduce the need for many-body interactions relative to isotropic models
%Since oriented pair potentials can capture features like bond bending and twisting that would require many-body interactions~\cite{Liwo2001, Larini2010} in isotropic models, using oriented pair potentials can reduce the need for many-body terms.

%Applying our scheme to proteins, we develop mapping and backmapping functions for a coarse-grained Protein model with Few AnisoTrop Sites (P-FAST).  

%Our paper is organized in three sections.  In Section~\ref{accuracy} we discuss the benefit of accurate coarse-grained models.  In Section~\ref{model} we introduce mapping and backmapping schemes for our new class of coarse-grained models, focusing on the specific case of proteins.  In Section~\ref{roundtrip} we discuss the metric of accuracy, apply it to our protein model, and compare our model's results with those of other coarse-grained protein models.  We conclude in section~\ref{conclusion}.

\section{Mapping and backmapping functions}
\label{mapping}

Mapping and backmapping functions are essential to coarse-grained and multiscale modeling.  Coarse-grained models consist of two parts, mapping functions and effective interactions.  The set of forward mapping functions $\textbf{M}$ defines the set of coarse-grained coordinates $\textbf{R}$ via $\textbf{R}=\textbf{M}(\textbf{r})$, where $\textbf{r}$ is the set of atomic positions.  In the next paragraph we will discuss how backmapping functions $\textbf{B}(\textbf{R})=\textbf{r}_{\rm back}$ allow coarse-grained models to 
approximately calculate even those observables
%\textit{approximately} predict observables 
that depend explicitly on the atomic positions $\textbf{r}$.  First, it is instructive to note that coarse-grained models can in principle be used to exactly calculate any observable that depends only on the reduced coordinates $\textbf{R}$.  For equilibrium observables, this can be seen by integrating out the variables eliminated by $\textbf{M}$ from the equilibrium distribution function~\cite{Saunders2013, Noid2013, Liwo2001}.  This results in an expression for the effective interactions $V(\textbf{R})$ that would ideally be applied to the coarse-grained model:
\begin{equation}
V(\textbf{R})=-k_{\textbf{B}}T\ln\left(\dfrac{\Omega}{\omega}\int d\textbf{r} \delta\left(\textbf{M}(\textbf{r})-\textbf{R}\right)\exp\left(-\beta v (\textbf{r})\right)\right).
\label{pmf}
\end{equation}
In Eq.~\ref{pmf}, $v(\textbf{r})$ is the all-atom potential energy function, and $\Omega$ and $\omega$ are the configurational volumes of the all-atom and coarse-grained systems.  The exact solution to Eq.~\ref{pmf} includes many-body terms of all orders, which would be prohibitively difficult to calculate and prohibitively slow to simulate.  Coarse-grained models are useful when Eq.~\ref{pmf} can be approximated by a small number of terms, e.g.
\begin{equation}
V(\textbf{R})\simeq \sum_{i=1}^{N_{\rm sites}}\sum_{j=1}^{i-1}V_{\rm pair}(\textbf{R}_i, \textbf{R}_j),
\label{pair}
\end{equation}
where $V_{\rm pair}$ is a pair interaction between sites $\textbf{R}_i$ and $\textbf{R}_j.$  
We will show that backmapping functions can be developed to predict atomic positions from the position and orientation of individual coarse-grained sites, $\vec{b}_j=\vec{B}_j({\bf R}_i)$.  Thus, many of the atomic-scale pair interactions represented by Eq.~\ref{pmf} should be expressible by coarse-grained pair potentials of the form of Eq.~\ref{pair}.  This will allow accurate effective interactions to be directly inverted from pair distribution functions calculated with small all-atom simulations, minimizing the need for fitting.
%As we will discuss in Section~\ref{interactions}, using oriented sites should allow for straightforward parameterization and implementation of accurate effective interactions following Eq.~\ref{pair}.  Even before having this discussion, it is clear from Eq.~\ref{pmf} that the forward mapping functions are crucial to coarse-grained models because they define both the sites and the target effective interactions.

Backmapping functions are equally crucial in their own right, either for predicting observables that depend explicitly on the atomic coordinates $\bf{r}$ or, as will be discussed in Section~\ref{multiscale}, for interfacing with all-atom simulations in multiscale schemes.
%\textit{Backmapping} functions are equally crucial  to any coarse-grained model that seeks to predict atomically precise observables either directly or via multiscale methods.  
Suppose that we want to use a coarse-grained model to calculate an observable $\mathcal{O}$ that depends explicitly on the atomic positions $\textbf{r}$.  For simplicity, assume that $\mathcal{O}$ depends linearly on $\textbf{r}$, $\mathcal{O}=\textbf{O}\cdot\textbf{r}$.  In addition to the error associated with approximating Eq.~\ref{pmf} by e.g.~Eq.~\ref{pair}, we expect an error resulting from our inability to precisely know the positions $\textbf{r}$.  We can only calculate $\mathcal{O}$ via a set of backmapping functions $\textbf{B}$ defining backmapped atomic positions $\textbf{r}_{\rm back}=\textbf{B}(\textbf{R})$.  The error associated with this backmapping (beyond any error associated with the effective interactions) can be calculated by defining the roundtrip root-mean-square dispacement (rmsd)
\begin{equation}
\Delta_{\textbf{r}}=\left(\left\langle\left(\textbf{r}_{\rm back}-\textbf{r}\right)^2\right\rangle\right)^{1/2}=\left(\left\langle\left(\textbf{B}\left(\textbf{M}(\textbf{r})\right)-\textbf{r}\right)^2\right\rangle\right)^{1/2},
\label{deltar}
\end{equation}
where the average is over the equilibrium distribution of the atomistic system.  The error in $\mathcal{O}$ depends on the covariance matrix for $\textbf{r}$, $\Sigma_{\textbf{r}}$, via $\Delta\mathcal{O}=\left(\textbf{O}\cdot\Sigma_{\textbf{r}}\cdot\textbf{O}^\intercal\right)^{1/2}$, and the elements of $\Sigma_{\textbf{r}}$ are bounded by the elements of $\Delta_{\textbf{r}}$ by the Pearson correlation coefficient equation.  Thus, to reduce the error in $\mathcal{O}$ associated with backmapping, we want to reduce the roundtrip rmsd of all the atomic positions.  
%This assumes that effective interactions can be found that are consistent with the backmapping functions, a point that will be discussed in Section~\ref{interactions}.  As discussed in Section~\ref{multiscale}, reducing $\Delta_{\textbf{r}}$ is equally important for multiscale schemes that use coarse-grained models to accelerate equilibration or sampling of all-atom simulations rather than predicting observables directly.

\section{Method}
\label{method}

Our method can be described generally as follows.  We defined a set of forward mapping functions $\textbf{M}(\textbf{r})=\textbf{R}$ that would integrate out the stiff degrees of freedom found in organic chemistry (bond lengths and angles, double-bond dihedral angles, and dihedral angles in rings) while retaining a nearly 1:1 correspondence between $\textbf{M}$ and the soft degrees of freedom (non-ring, single-bond dihedral angles).  Then, we defined backmapping functions $\textbf{B}(\textbf{R})=\textbf{r}_{\rm back}$ that would take advantage of this 1:1 correspondence to back out atomic coordinates with a precision limited only by fluctuations in the stiff degrees of freedom.  Finally, we numerically calculated optimal coefficients for the backmapping functions that minimized the root-mean-square displacement between initial and backmapped atomic positions ($\Delta_{\textbf{r}}$ in Eq.~\ref{deltar}), when applying the roundtrip mapping to a set of atomic configurations.

This procedure could be applied to any set of atomic configurations.  The formal connection to a potential of mean force requires that the set be an equilibrium distribution of configurations.  However, since there are no such experimental distributions with sufficient statistics to perform the optimization of $\textbf{B}$, and since creating such a distribution with all-atom simulations would be computationally expensive, we instead used the Protein Data Bank (PDB)~\cite{PDB} as a large, convenient, and experimentally based proxy for an equilibrium distribution.  In a general sense this is analogous to using the PDB to construct knowledge-based potentials for coarse-grained models~\cite{Li2013}.  However, in contrast to these approaches, our use of the PDB as a proxy only assumes that the local degrees of freedom integrated out by $\textbf{M}$ are in equilibrium in the PDB, making no assumptions about non-local degrees of freedom like those responsible for protein folding.

Details about how we extracted data from the PDB, dealt with missing or incorrect data, and treated indistinguishable atoms appear in the Supporting Information.  The Supporting Information also contains the optimal parameters for the backmapping functions, a list of rmsds by amino acid and atom type, and the C code we used to optimize the backmapping functions, calculate the rmsds, and generate the molecular files and TCL scripts used to create the \href{http://www.ks.uiuc.edu/Research/vmd/}{VMD}~\cite{VMD} images in this paper.

\section{Results}
\label{results}

\begin{figure*}
\includegraphics[width=\textwidth]{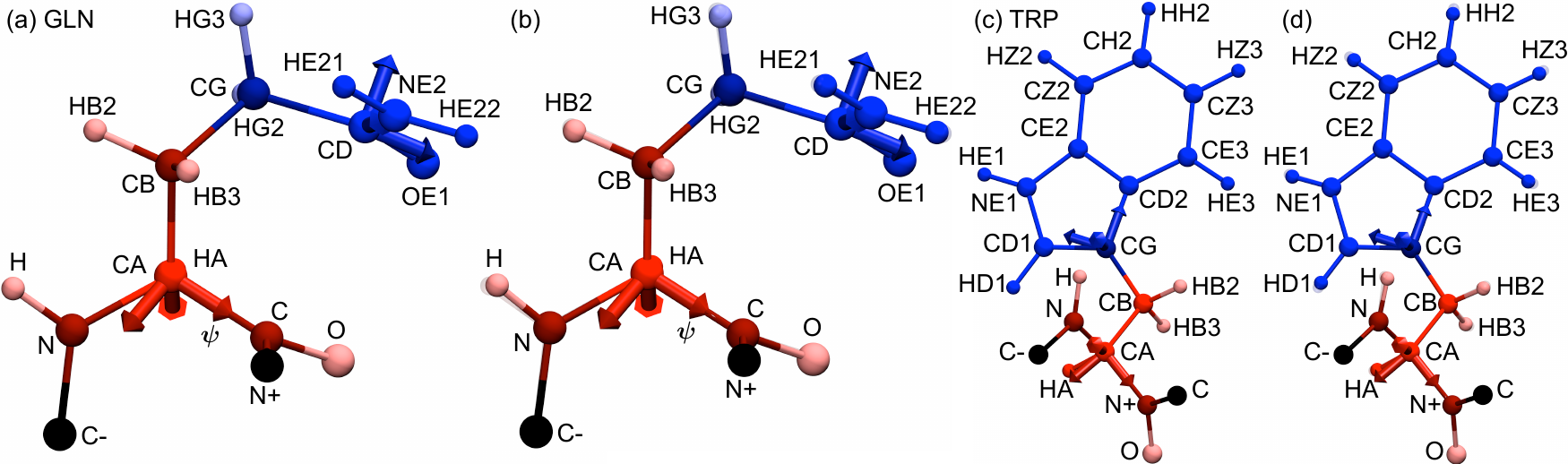}
\caption{Illustration of forward mapping and backmapping functions for non-terminal glutamine and tryptophan residues, using example residues from the PDB structure 2iqx of malate synthase G~\cite{Grishaev2008}.  Atoms are labeled using standard PDB atom names.  Examples for the other eighteen amino acids appear in Fig.~\ref{hydroxyl} and the Supporting Information.  (a) Glutamine atoms from the PDB structure (spheres) are mapped to two oriented sites (vectors).  The backbone site (red) is defined by the positions of the atoms CA, C, and N, and the sidechain site (blue) is defined by the atoms CD, NE2, and OE1.  Atoms are colored according to to which site they are associated with.  Neighboring atoms in adjacent residues are colored black.  (b) Backmapped atom positions (opaque) superposed on original positions (semi-transparent).  Lighter-colored atoms (e.g. HB2 and HB3) are backmapped using a linear correction that depends on the backmapped position of the nearest atom in an adjacent site (e.g. CG), which is colored darker.  (c) A tryptophan residue from 2iqx is mapped onto two oriented sites.  The backbone site (red) is defined by the positions of the atoms CA, C, and N, and the sidechain site (blue) is defined by the atoms CG, ND2, and CD1.  (d) Backmapped atom positions (opaque) superposed on original positions, as in (b).
%Atoms are colored red (backbone), blue, and green according to their primary associated site, and the orthonormal basis vectors representing each site are overlaid in the same color.  Lighter colored atoms 
%(a-b) Illustration of the mapping and backmapping schemes for an arginine residue (residue ID 9) from a PDB structure (ID 2iqx) of malate synthase G~\cite{PDB, Grishaev2008}.  Atoms are colored red (backbone), blue, and green according to their associated site, and the orthnormal basis vectors of each site 
}
\label{schematic}
\end{figure*}

Figure~\ref{schematic} illustrates how we constructed forward mapping and backmapping functions to take advantage of the stiff degrees of freedom found in proteins.  As shown in Fig.~\ref{schematic} (a) for one non-terminal glutamine residue drawn from the PDB, we grouped each amino acid residue into one to three coarse-grained sites $\textbf{r}_i$, each defined by a position and three orthonormal vectors, $\textbf{R}_i=\{\vec{R}_i, \hat{E}_{ix}, \hat{E}_{iy}, \hat{E}_{iz}\}$.  We defined the forward mapping function $\textbf{M}_i$ for each site $i$ as a function of the positions of three atoms, $\vec{r}_{\nu_{i1}}$, $\vec{r}_{\nu_{i2}}$, and $\vec{r}_{\nu_{i3}}$.  ($\nu_{in}$ defines the index of the $n$th atom defining the $i$th site.)  We let $\vec{r}_{\nu_{i1}}$ define the site position, $\vec{R}_i=\vec{r}_{\nu_{i1}}$, and we let the other two sites define the orthonomal vectors,
\begin{equation}
\begin{array}{l}
\hat{E}_{ix}=\hat{N}\left(\vec{r}_{\nu_{i1}}-\vec{r}_{\nu_{i0}}\right),\\
\hat{E}_{iz}=\hat{N}\left(\left(\vec{r}_{\nu_{i1}}-\vec{r}_{\nu_{i0}}\right)\times\left(\vec{r}_{\nu_{i2}}-\vec{r}_{\nu_{i0}}\right)\right),\\
\hat{E}_{iy}=\hat{N}(\hat{E}_{iz}\times\hat{E}_{ix}),
\end{array}
\end{equation}
where $\hat{N}(\vec{r})=\vec{r}/|\vec{r}|$.
For example, we placed a backbone site (red arrows in Fig.~\ref{schematic} (a)) at each $\alpha$-carbon (CA) and defined the backbone site orientation by the positions of the neighboring carbonyl carbon (C) and nitrogen (N) atoms in the backbone.  Since bond lengths and angles tend to be distributed close to their average values, we could immediately write down backmapping functions for all atoms directly bound to the central atom.  For an atom $j$ of type $\alpha$ (e.g. $\beta$-carbons in glutamine) belonging to site $i$, we simply defined the backmapping function $\vec{B}_\alpha(\textbf{R}_i)$ via
\begin{equation}
\vec{B}_\alpha(\textbf{R}_i)=\vec{R}_i+c_{\alpha x}\hat{E}_{ix}+c_{\alpha y}\hat{E}_{iy}+c_{\alpha z}\hat{E}_{iz},
\label{directbackmap}
\end{equation}
where $\vec{c}_{\alpha}$ is the average position of atoms of type $\alpha$ in the frame of their coarse-grained site.  With this backmapping, the roundtrip rmsd becomes simply the rmsd of the atom positions in the frame of the site.  As illustrated by the close agreement between backmapped (solid) and original (semi-transparent) atom positions in Fig.~\ref{schematic} (b), all atoms bound to the $\alpha$-carbon tend to move only small distances during the roundtrip mapping.  Averaging over all residues in the PDB, we find that $\alpha$-carbons, carbonyl carbons, nitrogens, $\beta$-carbons (CB), and $\alpha$-hydrogens (HA) move on average 0, 0.01, 0.07, 0.08, and 0.06~\AA, respectively.  We can account for the model's ability to backmap five atom positions (15 degrees of freedom) from one site (six degrees of freedom) because the tetrahedral cluster is constrained by four bond lengths and five independent bond angles, each largely conserved across the PDB.

\begin{table*}
{\small
\begin{filecontents*}{site_table.csv}
Res,Atom 1,Atom 2,Atom 3,Direct atoms,Corrected atoms
ALA, CA , C  , N  ,HA CB HB2 HB3 HB1,O H
ARG, CA , C  , N  ,HA,O H
ARG, CG , CD , CB ,HG2 HG3,HB2 HB3 HD2 HD3
ARG, CZ , NH1, NH2,NE HH11 HH12 HH21 HH22,HE
ASN, CA , C  , N  ,HA CB,O H HB2 HB3
ASN, CG , ND2, OD1,HD21 HD22,
ASP, CA , C  , N  ,HA CB,O H HB2 HB3
ASP, CG , OD2, OD1,HD2,
CYS, CA , C  , N  ,HA,O H
CYS, SG , CB , HG ,,HB2 HB3
GLN, CA , C  , N  ,HA CB,O H HB2 HB3
GLN, CD , NE2, OE1,CG HE21 HE22,HG2 HG3
GLU, CA , C  , N  ,HA CB,O H HB2 HB3
GLU, CD , OE2, OE1,CG HE2,HG2 HG3
GLY, CA , C  , N  ,HA2 HA3,O H
HIS, CA , C  , N  ,HA CB,O H HB2 HB3
HIS, CG , ND1, CD2,CE1 NE2 HD1 HD2 HE1 HE2,
ILE, CA , C  , N  ,HA,O H
ILE, CG1, CB , CD1,HG12 HG13 HD11 HD12 HD13,CG2 HG21 HG22 HG23 HB
LEU, CA , C  , N  ,HA CB,O H HB2 HB3
LEU, CG , CD1, CD2,HG HD11 HD12 HD13 HD21 HD22 HD23,
LYS, CA , C  , N  ,HA,O H
LYS, CG , CB , CD ,HG2 HG3,HB2 HB3
LYS, CE , NZ , CD ,HE2 HE3 HZ1 HZ2 HZ3,HD2 HD3
MET, CA , C  , N  ,HA CB,O H HB2 HB3
MET, SD , CE , CG ,HE1 HE2 HE3,HG2 HG3
PHE, CA , C  , N  ,HA CB,O H HB2 HB3
PHE, CG , CD1, CD2,CE1 CE2 CZ HD1 HD2 HE1 HE2 HZ,
PRO, CA , C  , N  ,HA,O H
PRO, CG , CD , CB ,HG2 HG3,HB2 HB3 HD2 HD3
SER, CA , C  , N  ,HA,O H
SER, OG , CB , HG ,,HB2 HB3
THR, CA , C  , N  ,HA,O H
THR, OG1, CB , HG1,,HB CG2 HG21 HG22 HG23
TRP, CA , C  , N  ,HA CB,O H HB2 HB3
TRP, CG , CD2, CD1,NE1 CE2 CE3 CZ2 CZ3 CH2 HD1 HE1 HE3 HZ2 HZ3 HH2,
TYR, CA , C  , N  ,HA CB,O H HB3 HB2
TYR, CG , CD1, CD2,CE1 CE2 CZ OH HD1 HD2 HE1 HE2 HH,
VAL, CA , C  , N  ,HA,O H
VAL, CB , CG1, CG2,HB HG11 HG12 HG13 HG21 HG22 HG23,
\end{filecontents*}
\csvautotabular{site_table.csv}}
\label{rmsd}
\caption{List of atoms defining the forward mapping and backmapping functions for each site.  The first column lists the residue, the next three colums list the atoms $\nu_{ij}$, $j=1$, 2, 3, defining the the forward mapping function $\textbf{M}_i$.  The fifth column lists additional atoms backmapped directly by Eq.~\ref{directbackmap}.  The last column lists the atoms backmapped by Eq.~\ref{lincorrection}.}
\label{sitetable}
\end{table*}

\begin{figure*}
\includegraphics[width=\textwidth]{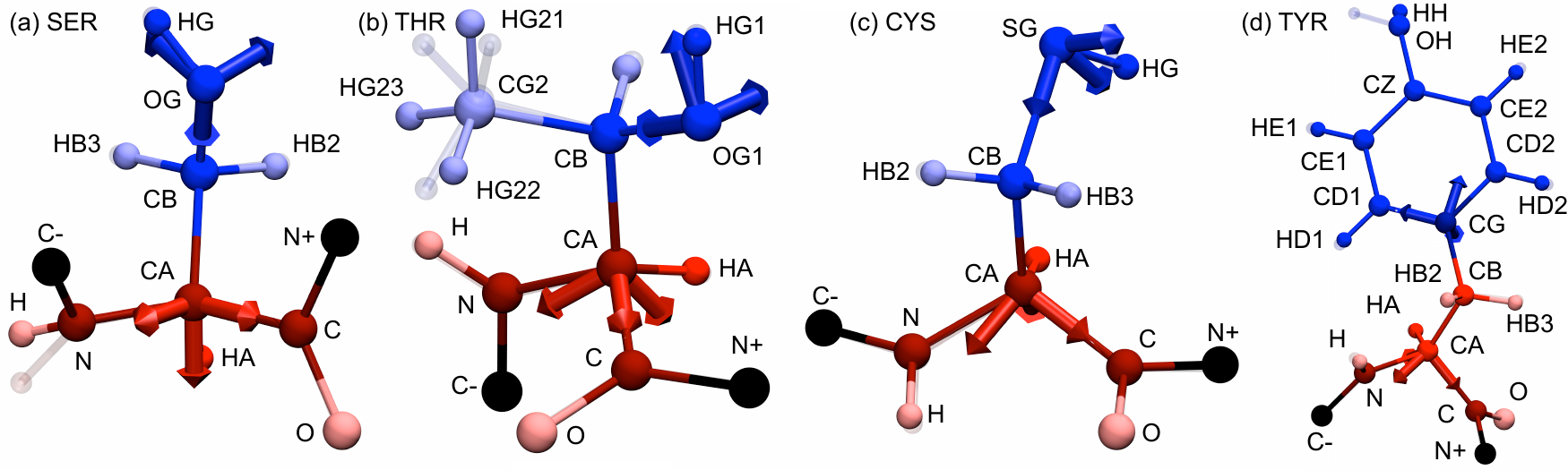}
\caption{Backmapped atom positions (opaque) superposed on original positions (semi-transparent) for (a) serene, (b) threonine, (c) cysteine, and (d) tyrosine.  In (a-c) the hydroxyl or thiol groups help define the sidechain site.  In (d) the torsion of the hydroxyl group is unconstrained, leading to a large disagreement between the original and backmapped HH position. 
}
\label{hydroxyl}
\end{figure*}

We found that directly backmapping backbone atoms separated by two bonds from the $\alpha$-carbon did not result in low roundtrip rmsds, because the positions of these atoms in the frame of the backbone site each depend on a soft torsional degrees of freedom.  For example, the position of the carbonyl oxygen (O) in the $i$th residue depends on the $\psi$ dihedral angle (torsion of the CA-C bond, see Fig.~\ref{schematic} (a)).  Fortunately, the same torsional degree of freedom controls the position of the nitrogen in the $(i+1)$th residue, denoted $N^+$ in Fig.~\ref{schematic}.  Since the position of the $(i+1)$th nitrogen is predicted accurately by direct backmapping from the $(i+1)$th backbone site, we could predict the position of the $i$th oxygen by fitting it to a linear function of the predicted position of the $(i+1)$th nitrogen.  All of this is done in the frame of the $i$th backbone site.  A linear function is sufficient because the dihedral angle rotation relating the oxygen and nitrogen positions is a linear transformation.  In general, for an atom $l$ of type $\beta$ directly backmapped by site $k$, we define $\vec{B}_\alpha^\prime(\textbf{R}_{i}, \textbf{R}_{k})$ as its predicted position in the frame of site $i$,
\begin{equation}
\vec{B}_\beta^\prime(\textbf{R}_{i}, \textbf{R}_{k})=\left(\vec{B}_\beta(\textbf{R}_{k})-\vec{R}_i\right)\cdot\{\vec{E}_{ix}, \vec{E}_{iy}, \vec{E}_{iz}\}.
\end{equation}
Then, if $j$ is the atom of type $\alpha$ in site $i$ whose position is correlated with the position of atom $l$ of type $\beta$, the backmapping function for $j$ with a linear correction is
\begin{equation}
\vec{B}_\alpha(\textbf{R}_i, \textbf{R}_k)=\vec{R}_i+c_{\alpha \beta x}^{~\prime}\hat{E}_{ix}+c_{\alpha \beta y}^{~\prime}\hat{E}_{iy}+c_{\alpha \beta z}^{~\prime}\hat{E}_{iz},
\label{lincorrection}
\end{equation}
where
\begin{equation}
%\vec{c}_{ijkl}^{~\prime}=\vec{c}_{ij}+\vec I_{ikjl} + S_{ijkl}\cdot \vec{B}_l^\prime(\textbf{R}_{i}, \textbf{R}_{k}).
\vec{c}_{\alpha \beta}^{~\prime}=\vec I_{\alpha \beta} + S_{\alpha \beta}\cdot \vec{B}_\beta^\prime(\textbf{R}_{i}, \textbf{R}_{k}).
\end{equation}
We calculated the 3-vector $\vec I_{\alpha \beta}$ and the $3\times 3$ matrix $S_{\alpha \beta}$ by analytically performing ordinary least-squares fits to the three components of the vector equation
\begin{equation}
\left\langle\left(\vec{r}_j-\vec{R}_{i}\right)\cdot\{\vec{E}_{ix}, \vec{E}_{iy}, \vec{E}_{iz}\}\right\rangle_{j\in \alpha}=\vec{c}_{\alpha\beta}^{~\prime},
\end{equation}
where the average is over all atoms $j$ of type $\alpha$.

The three-bond spacing between $\alpha$-carbons in proteins is ideal for backmapping accurate atomic positions from oriented backbone sites.  For each residue we backmapped the C, N, HA, and CB atoms directly via Eq.~\ref{directbackmap}, and we backmapped the O and H atoms via Eq.~\ref{lincorrection} using the predicted positions of the N$^+$ and C$^-$ atoms, respectively.  As illustrated by the close agreement between the solid and semi-transparent O and H atoms in Fig.~\ref{schematic} (b), the linear correction resulted in small displacements during the roundtrip mapping.  Averaged over the PDB, O atoms moved 0.086~\AA~and H atoms moved 0.274~\AA.  If residues were separated by more than three bonds, such an accurate backmapping would not be possible with only one backbone site per residue, because additional soft torsional degrees of freedom would be unconstrained.

Moving up the main branch of each amino acid sidechain, we added as few additional sites as possible while ensuring that sites were spaced by no more than three bonds.  Columns 2-4 of Table~\ref{sitetable} lists the atoms defining each site, using standard atom names from the PDB.  As for the backbone sites, we used Eq.~\ref{directbackmap} to directly backmap those atoms whose relative positions could be constrained by stiff degrees of freedom (Column 5 of Table~\ref{sitetable}), and we used linear corrections to backmap those atoms off the main chain that required a torsional degree of freedom to be constrained (last column of Table~\ref{sitetable}).  When correcting the position of an atom, we always corrected it using the predicted position of the closest main-chain atom on a neighboring site.  For example, for glutamine (Fig.~\ref{schematic}(a)) we corrected the $\beta$-hydrogens in the backbone site (HB2 and HB3) by the predicted position of the $\gamma$-carbon in the sidechain site (CG), and we corrected the $\gamma$-hydrogens in the sidechain site (HG2 and HG3) by the predicted position of the $\beta$-carbon in the backbone site (CG).  The close agreement between solid and semi-transparent HB2, HB3, HG2, and HG3 atoms in Fig.~\ref{schematic} (b) illustrates the high accuracy of these roundtrip mappings.

Although the sites in some sidechains had to spaced by fewer than three bonds, 
%Although our spacing of some sidechain sites by less than three bonds made imperfect use of some of the stiff degrees of freedom, 
using oriented sites more than made up for this redundancy by (1) utilizing the stiff torsional degrees of freedom in the rings of histidine, phenylalanine, tryptophan, and tyrosine and (2) allowing accurate estimation of methyl and amine hydrogens with no additional overhead.  Tryptophan's double ring is a dramatic example of the first case.  As shown in Fig.~\ref{schematic} (c), we placed tryptophan's sidechain site on the $\gamma$-carbon (CG), only two bonds away from the $\alpha$-carbon (CA) of the backbone site.  This meant that the $\beta$-carbon (CB) could be predicted by either site; we chose to associate it with the backbone.  Although this meant that the sidechain site's orientation was not used to predict the position of the $\beta$-carbon, the stiffness of the double ring meant that we could use the site's orientation to predict the position of \textit{twelve} atoms not directly bonded to the $\gamma$-carbon.  As illustrated in Fig.~\ref{schematic} (d), the direct backmapping of these atoms resulted in low roundtrip rmsds.  Averaging over the PDB, the largest roundtrip displacement in tryptophan's double ring was for HZ3, which moved on average 0.13~\AA.  Our ability to model ring groups using few degrees of freedom starkly contrasts the ability of point-size coarse-grained models that typically require more sites per atom to model rings~\cite{Marrink2007, Monticelli2008}.

%methyl: ALA, ILE, LEU, MET, VAL
%amine: ARG, GLN, neutral N-terminus
%charged amine: LYS, charged N-terminus

Glutamine's amine group (Fig.~\ref{schematic} (c)) is an example of the second case, where methyl or amine hydrogens could be backmapped with high accuracy and no additional overhead.  When adding sites, we ensured that methyl carbons and amine nitrogens were within one bond of a coarse-grained site, but we imposed no such requirement for the methyl and amine hydrogens.  As a result, the hydrogens' positions in the frame of the site depend on the torsion of the bond connecting the carbon or nitrogen to the site center.  However, we found that these hydrogen positions could be backmapped directly via Eq.~\ref{directbackmap} with relatively low rmsds of 0.11~\AA~for neutral amine hydrogens in arginine, asparagine, and glutamine; between 0.17~\AA~and 0.24~\AA~for methyl hydrogens in alanine, isoleucine, leucine, methionine, threonine, and valine; and 0.25~\AA~for quaternary ammonium hydrogens in lysine.  The low rmsds for the neutral amine hydrogens is due to the planarity of the guanidinium and amide groups.  The rmsds of the methyl and NH3+ groups are kept reasonably low from a combination of two factors.  First, we named the indistinguishable hydrogen atoms according to their dihedral angles, which imposes an upper bound of around 0.51~\AA~for methyl groups and 0.52~\AA~for NH3+ groups (see Supporting Information).  Second, these groups do show some preference for certain dihedral angles, reducing the rmsd to less than half of these upper bounds.
%ALA methyl 0.18
%ILE methyl 0.18, 0.19
%LEU methyl 0.17, 0.18
%MET methyl 0.24
%THR methyl (corrected) 0.20
%VAL methyl 0.24
%ARG amine 0.11
%ASN amine 0.11
%GLN amine 0.11
%LYS quaternary ammonium group 0.25, out 0.87
%As discussed in the Supporting Information, these rmsds are between 2.0 (quaternary ammonium) and 6.9 times smaller than what we would expect if the dihedral angles of these groups were uniformly distributed.
%ARG 0.11/
%ASN
%GLN

%To some extent, we might expect low rmsds simply because we resolved indistinguishable hydrogens according to their dihedral angle (see Supporting Information).  However, a simple calculation shows that this effect is not enough to constrain the amine and methyl hydrogens to such low rmsds.  We take lysine's NH3+ group as the most stringent example, since its hydrogens have the largest rmsd.  If we assume that the first hydrogen lies at a radial distance $r$ from the C-N axis and is uniformly distributed at a dihedral angle between 0 and $2\pi/3$, then its average radial position would be $r_0=(3\sqrt{3}/2\pi)r=0.83r$ and its rmsd would be $\sigma=(1-27/4\pi^2)^{1/2}r=0.56r$, with a ratio $r_0/\sigma=(4\pi^2/27-1)^{1/2}=0.68$.  In contrast, lysine's NH3+ hydrogens have $\sigma=0.25$~\AA~and $r_0=0.87$~\AA, with a ratio $r_0/\sigma=0.29$.

In contrast to the amine and methyl hydrogens, we found that hydroxyl and thiol groups in serene, threonine, tyrosine, and cysteine have broad distributions of dihedral angles, 
%controlling the positions of their terminal hydrogens, 
and the single terminal hydrogens in these groups have no indistinguishable partners that could be used to reduce their rmsd.  For serene, threonine, and cysteine, we were able to include the hydrogen as one of the three atoms defining the site without increasing the number of sites or adversely affecting the backmapping of other atoms, as shown in Fig.~\ref{hydroxyl} (a-c).   For tyrosine, the hydroxyl group lies on the opposite side of an aromatic ring from the site used to backmap the ring.  We chose to leave the hydroxyl torsion unconstrained in the backmapping, leading to an unusually large rmsd of 0.836~\AA~for the hydroxyl hydrogen atom (HH).  Since introducing another site (six degrees of freedom) to constrain this one degree of freedom would be inefficient, the best way to reduce the rmsd might be to introduce an additional internal degree of freedom to tyrosine's sidechain site.  This degree of freedom would control the backmapping of HH and modulate the site's effective interactions.

\begin{figure}
\includegraphics[width=0.5\textwidth]{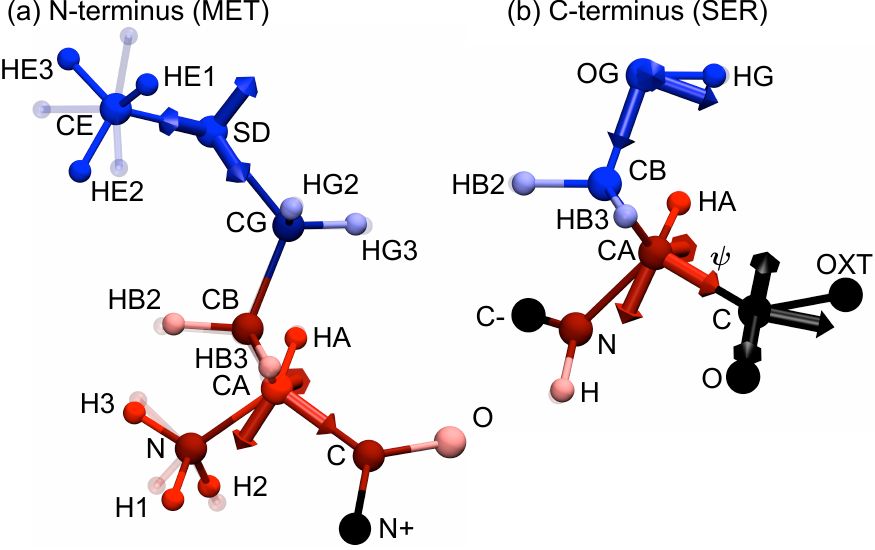}
\caption{Backmapped atom positions (opaque) superposed on original positions (semi-transparent) for (a) the N-terminal methionine and (b) the C-terminal serene from the PDB structure 2eyz of the CRK protein~\cite{Kobashigawa2007}.  In (a) the N-terminal quaternary ammonium hydrogens are backmapped from the position and orientation of the C$_\alpha$ site via Eq.~\ref{directbackmap}.  In (b) an additional C-terminal site is defined from the positions of atoms C, O, and OXT.
}
\label{termini}
\end{figure}

We modeled the N-terminal hydrogens similarly to how we modeled the amine hydrogens on the sidechains, backmapping their positions directly via Eq.~\ref{directbackmap} using the terminal backbone site, as shown in Fig.~\ref{termini} (a).  Since an appreciable number (2.4\%) of N-terminal amines in the PDB were neutral, we treated NH2 and HN3$^+$ groups separately.  Although these terminal amine groups occupy a broad distribution of dihedral angles, we were able to reduce the rmsd by naming the indistinguishable hydrogens according to their dihedral angles, bringing the rmsds down to 0.451~\AA~for NH2 hydrogens and 0.637~\AA~for NH3 hydrogens (see Supporting Information).  As for tyrosine's hydroxyl group, introducing a single internal degree of freedom to describe this torsion could significantly reduce these rmsds.

As shown in Fig.~\ref{termini} (b), the positions of the heavy atoms O and OXT at the C-terminus depend on the torsion of the C-terminal CA-C bond ($\psi$ dihedral angle).  To allow accurate backmapping, we introduced an additional site at the C-terminus defined by the positions of the atoms C, O, and OXT (black arrows in Fig.~\ref{termini} (b)).  Even with this additional site, the rmsd for the OXT atom is still moderately large, 0.347~\AA, mostly due to an unusually broad distribution for the C-OXT bond length in the PDB, $r_{\rm C-OXT}=1.249\pm0.327$~\AA.

We designed our model to target proteins in their most common charged states, but we also applied our roundtrip mapping to structures containing protonated carboxyl groups at the C-terminus and/or in aspartic acid or glutamic acid sidechains, which represented fewer than 1\% of the 2\% of PDB structures containing hydrogens.  Since the positions of the hydrogens in these groups depend on an unconstrained torsional degree of freedom, they had predictably large rmsds of 0.973~\AA, 0.809~\AA, and 0.804~\AA, respectively.  For the C-terminus and aspartic acid, these hydrogens could be constrained by using an alternate mapping for charged groups.  For example, changing the atoms defining the C-terminus site from C, O, and OXT (see Fig.~\ref{termini} (b)) to OXT, HXT, and C would constrain HXT up to variations in bond lengths and angles, while allowing O to be backmapped through Eq.~\ref{lincorrection}.  The main chain of the glutamic acid sidechain is too long for the additional hydrogen to be accurately backmapped without adding an additional degree of freedom.
%HXT: 18 compared to 2220 N-terminal hydrogen groups; 0.97323543
%(ARG: NH group only when neutral)
%ASP: 274 compared to 47000 HB's; 0.80935253
%GLU: 286 compared to 52500 HG's; 0.80439487

\begin{table}
{\small
\begin{filecontents*}{residue_table_nodiscard.csv}
Residue,Atoms,Sites,Mapping,Heavy rmsd,H rmsd
ALA,10,1,5.0,0.056,0.178
ARG,24,3,4.0,0.043,0.120
ASN,14,2,3.5,0.054,0.158
ASP,12,2,3.0,0.053,0.190
CYS,11,2,2.8,0.058,0.163
GLN,17,2,4.2,0.050,0.141
GLU,15,2,3.8,0.052,0.160
GLY,7,1,3.5,0.059,0.358
HIS,18,2,4.5,0.050,0.140
ILE,19,2,4.8,0.050,0.160
LEU,19,2,4.8,0.049,0.159
LYS,22,3,3.7,0.046,0.163
MET,17,2,4.2,0.078,0.191
PHE,20,2,5.0,0.048,0.135
PRO,15,2,3.8,0.050,0.063
SER,11,2,2.8,0.057,0.131
THR,14,2,3.5,0.057,0.168
TRP,24,2,6.0,0.050,0.134
TYR,21,2,5.2,0.053,0.334
VAL,16,2,4.0,0.045,0.209
Total, , , ,0.051,0.179
\end{filecontents*}
\csvautotabular{residue_table_nodiscard.csv}}
\caption{List of roundtrip mapping ratios and roundtrip rmsds by amino acid (top 20 rows) and overall (bottom row).  The mapping ratio $R$ is the ratio of the number of atomic degrees of freedom to the number of coarse-grained degrees of freedom, $R=N_{\rm atoms}/(2\times N_{\rm sites}.$  The rmsds are separated for heavy atoms and hydrogens.}
\label{rmsd}
\end{table}

\begin{figure*}
\includegraphics[width=\textwidth]{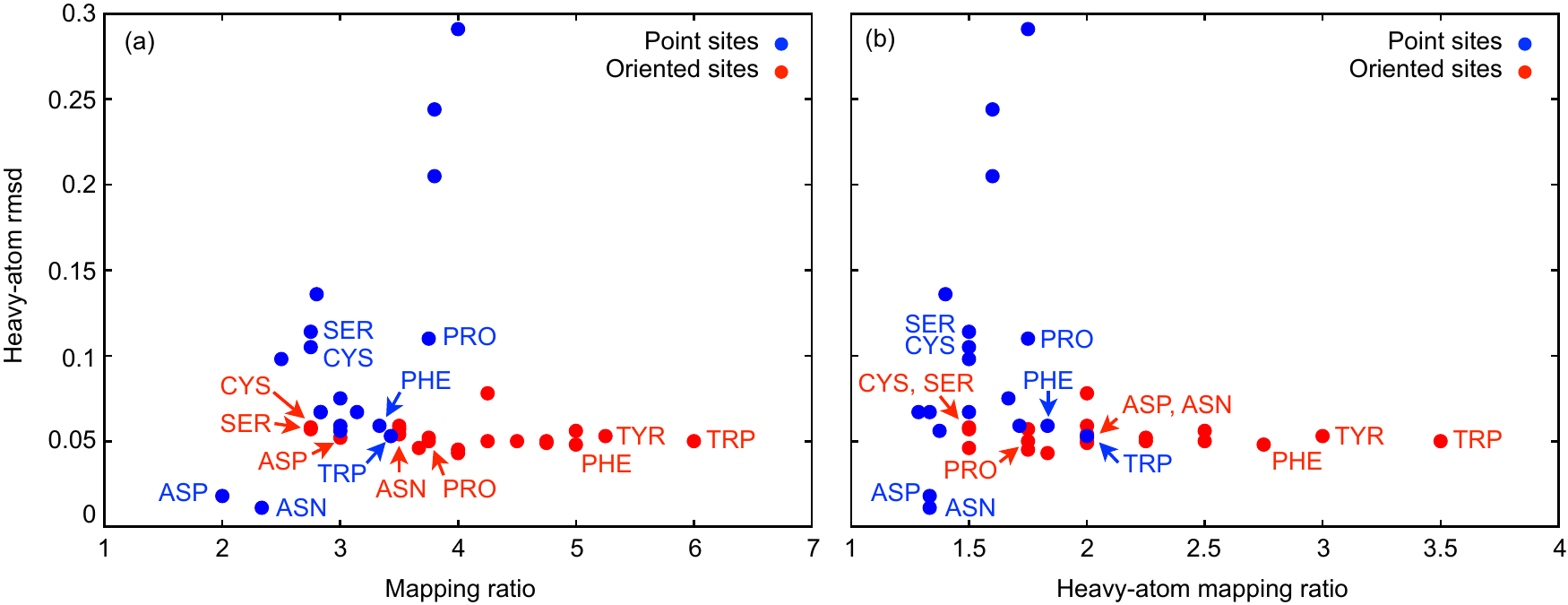}
\caption{Comparison of heavy-atom rmsds and mapping ratios for our model (red points) and an ``intermediate-resolution" point-site model (PRIMO~\cite{Gopal2009, Kar2013}) specifically designed for accurate backmapping (blue points).  Each point represents a different amino acid.  Panel (a) defines the mapping ratio as the number of atomic degrees of freedom per course-grained degree of freedom.  Panel (b) defines the mapping ratio as the number of non-hydrogen atomic degrees of freedom per coarse-grained degree of freedom.  Results for the point-site model are taken from Ref.~\cite{Gopal2009} and were obtained by performing a roundtrip mapping on 611 protein structures minimized with an all-atom forcefield.  In general, our model is both more accurate (smaller rmsds) and more efficient (larger mapping ratios).
}
\label{comparison}
\end{figure*}

Overall, we found that using oriented coarse-grained sites allowed us to efficiently store the information necessary to backmap accurate all-atom configurations.  Averaged over the PDB, heavy atoms moved \heavyrmsd~\AA~during our roundtrip mapping, while hydrogens moved \hrmsd~\AA.  Table~\ref{rmsd} lists the mapping ratios and rmsd for each amino acid.  A complete list broken down by atom type appears in the Supporting Information.  The low rmsds are remarkable given that the average mapping ratio is 4.1; that is, information associated with 76\% of the atomic degrees of freedom are discarded during the roundtrip mapping.

To get a sense of how much of this efficiency can be attributed to using oriented sites, in Fig.~\ref{comparison} (a) we compare heavy-atom rmsds and mapping ratios for our model with those calculated in Ref.~\cite{Gopal2009} for an ``intermediate-resolution" point-site model (PRIMO) 
which was designed for accurate backmapping~\cite{Gopal2009}.
PRIMO's high resolution~\cite{Gopal2009} and detailed effective interactions~\cite{Kar2013} has allowed it model folded, folding~\cite{Kar2013}, and membrane~\cite{Kar2014} proteins with higher accuracy than typical for coarse-grained models.
%and has been shown to accurately stabilize large proteins, fold small proteins, and accurately reproduce thermodynamic and structural features of membrane proteins~\cite{Kar2014}.
%specifically designed for accurate backmapping~\cite{Gopal2009, Kar2013}.  
PRIMO maps each residue to between 4 and 8 structureless sites.  As a result, it reduces the number of atomic degrees of freedom by factors ranging between 2 (aspartic acid) and 4 (valine), as shown by the blue points in Fig.~\ref{comparison} (a),
% depending on residue (red points in Fig.~\ref{comparison} (a)), 
with an average mapping ratio of 3.0.  
This is a substantially finer mapping than most point-site coarse grained models.  The average mapping ratios for the UNRES~\cite{He2009}, MARTINI~\cite{Monticelli2008}, and MS-CG~\cite{Hills2010} models are 8.2, 7.1, and 6.0, while Spiga \textit{et al}.'s model including dipole degrees of freedom has an average mapping ratio of 6.6.
Ref.~\cite{Gopal2009} only performed a roundtrip mapping with PRIMO for heavy atoms (finding an overall rmsd of 0.099~\AA), so we only plot heavy-atom rmsds in Fig.~\ref{comparison}.  Since the calculation of the heavy-atom rmsds does not use the mapping functions for hydrogens, it is instructive to also compare the heavy-atom rmsds to \textit{heavy-atom} mapping ratios, the number of non-hydrogen atomic degrees of freedom per coarse-grained degree of freedom.  As shown by the blue points in Fig.~\ref{comparison} (b), PRIMO's heavy-atom mapping ratios fall between 1.3 (lysine) and 2 (tryptophan), with an average of 1.5

Comparing PRIMO to our model (red points in Fig.~\ref{comparison}), we find that across amino acids our model is more accurate (smaller rmsds) and more efficient (larger mapping ratios).  The worst cases for our model are cysteine and serine, whose sidechains are two heavy atoms long, requiring a sidechain site to backmap the position of only one heavy atom (and three hydrogens) that could not be directly backmapped from the backbone site (see Fig.~\ref{hydroxyl} (a) and (c)).  Our mapping ratios for these amino acids are no larger than PRIMO's (2.75, or 1.5 considering only heavy atoms), but our backmapping is more accurate (rmsds of 0.58 and 0.57~\AA~vs 0.105 and 0.114~\AA~for PRIMO).  Our proline mapping also has a ratio equal to its PRIMO counterpart (3.75, or 1.75 considering only heavy atoms) and a lower rmsd (0.050~\AA~vs 0.110~\AA).  The only residues with lower rmsds in PRIMO are asparagine, aspartic acid, and methionine (0.011, 0.018, and 0.067~\AA~vs 0.054, 0.053, and 0.078~\AA), but PRIMO achieves these low rmsds by using a nearly atomistic heavy-atom mapping ratio of 4/3, compared to a ratio of 2 used for these amino acids in our model.  The remaining fourteen amino acids have both larger mapping ratios and smaller rmsds in our model.  Amino acids with rings perform best in our model, due to the fact that the planarity of the rings allows the structure of an entire ring (or double ring) to be accurately stored in a single oriented site.  Histidine, phenylalanine, tyrosine, and tryptophan all have smaller rmsds than in PRIMO despite having mapping ratios that are at least 50\% larger.  

\section{Discussion}
\label{multiscale}

%By demonstrating that our model can accurately store atomic positions while discarding 76\% of the atomic degrees of freedom, we have justified the substantial computational task of parameterizing the effective interactions that would complete our model.  The extreme accuracy apparent in our backmapping functions will translate to accuracy and simplicity in our effective interactions.  Since the positions of most atoms can be backmapped directly from the positions and orientations of individual coarse-grained sites, any atomic-scale interaction depending on these atom positions should be captured by pairwise potentials of mean force.  (Interactions that depend sensitively on positions of atoms backmapped through the linear correction, Eq.~\ref{lincorrection}, may require some many-body terms.)  Although using all-atom simulations to calculate these potentials

%By demonstrating that our model can accurately store atomic positions while discarding 76\% of the atomic degrees of freedom, we know that equipping our model with similarly accurate effective interactions will allow 

By demonstrating that our model can store atomic positions with an unprecedented combination of accuracy (small rmsds) and efficiency (large mapping ratios),
we have justified the substantial computational task of parameterizing the effective interactions necessary to complete our model.  Calculating multidimensional potentials of mean force from all-atom simulations should allow us to write down effective interactions capturing atomic-scale interactions, because the relative atomic positions controlling these interactions are related directly to the relative positions and orientations of neighboring coarse-grained sites.  As discussed in Section~\ref{mapping}, equipping our model with these effective interactions should allow us to calculate any observable of the atomistic system with an error proportional to the small roundtrip rmsds calculated in this paper.

Having designed a model that preferentially integrates out the stiffest degrees of freedom (bond bending, bond stretching, and dihedral rotations in rings) we think that we have approached the accuracy limit of coarse-grained modeling.  Nevertheless, there are undoubtedly some molecular processes so sensitive to error that they could not be modeled even by an optimal coarse-grained model.  For such systems, coarse-grained models can be used to accelerate equilibration, sampling, or dynamics of all-atom simulations in various multiscale schemes~\cite{Tschop1998, Shih2007, Perlmutter2009, Neri2005, Shi2006, Machado2011, Mamonov2012, diPasquale2012, Leguebe2012, Lyman2006, Lyman2006b, Christen2006, Moritsugu2010}.

A small roundtrip rmsd is a clear figure of merit for a coarse-grained model's performance in all multiscale schemes.  Sequential multiscale modeling~\cite{Tschop1998, Shih2007, Perlmutter2009} requires that configurations be handed back and forth between coarse-grained and all-atom simulations without lengthy relaxations.  Reducing the roundtrip rmsd ensures that relaxations are short.  Embedded multiscale modeling~\cite{Neri2005, Shi2006, Machado2011, Mamonov2012, diPasquale2012, Leguebe2012} requires that coarse-grained and all-atom regions can be stitched together in the same simulation box.  When a coarse-grained can backmap accurate all-atom configurations, the coarse-grained region can be seamlessly blended into the all-atom region using the backmapped all-atom configurations.

Multiscale replica exchange simulations~\cite{Lyman2006, Lyman2006b, Christen2006, Moritsugu2010} 
%are the most rigorous multiscale approach, because they are designed so that the coarse-grained model 
require that coordinates be exchanged in equilibrium along a ladder of otherwise equivalent system ``replicas" differing in resolution.
These exchanges allow the higher replicas (coarse-grained simulations) to accelerate sampling of the lowest replica (the all-atom simulation) without biasing the all-atom simulation.
The efficiency of a multiscale replica exchange simulation is determined by the number of replicas and the exchange acceptance rates.  Using a high-resolution coarse-grained model in this approach would increase the overlap in distributions sampled by the highest and lowest replicas, thereby increasing acceptance rates and/or requiring fewer replicas.
\\

\section{Conclusion}
\label{conclusion}

We have demonstrated the accuracy of a coarse-grained protein model with oriented coarse-grained sites by calculating the distance atoms move during a roundtrip mapping.  By preferentially integrating out stiff degrees of freedom associated with bond stretching, bond bending, and bond twisting in rings, our model achieves a combination of accuracy (small rmsds) and efficiency (large mapping ratios) that has not previously been attained by coarse-grained protein models.  Our model's nearly atomistic resolution should allow for parameterization of detailed effective interactions accounting for atomic-scale interactions.  Once equipped with these effective interactions, our model should be able to extend the reach of time and length scales accessible to molecular simulation, either through direct simulations or through seamless integration into multiscale simulation schemes. 

\section{Acknowledgement}

We thank Ranjan Mannige, Steve Whitelam, and Ron Zuckermann for useful comments on the manuscript.  This project was funded by the Defense Threat Reduction Agency under Contract No. IACRO-B1144571.  Work at the Molecular Foundry was supported by the Office of Science, Office of Basic Energy Sciences, of the U.S. Department of Energy under Contract No. DE-AC02-05CH11231.

\section{Supporting information}

(1) Supporting document including (a) descriptions of how we extracted data from the PDB, dealt with missing or incorrect data, and treated indistinguishable atoms, (b) images illustrating the forward mapping and backmapping of the remaining amino acids, (c) list of roundtrip rmsds by amino acid and atom type, and (d) list of the calculated optimal parameters for the model.  (2) Source code in C that we used to optimize the backmapping functions, calculate the rmsd, and generate the molecular files and TCL scripts used to create the \href{http://www.ks.uiuc.edu/Research/vmd/}{VMD}~\cite{VMD} images in this paper.

\bibliography{protein.bbl}
%\bibliography{../../../Papers/Bibliography/Bibliography}

\end{document}